\documentclass[aps,prl,reprint,superscriptaddress,amsmath,amssymb,floatfix,nobalancelastpage]{revtex4-2}

\usepackage{graphicx}
\usepackage{bm}
\usepackage{xcolor}
\usepackage{microtype}
\usepackage[hypertexnames=false,colorlinks=true,allcolors=blue]{hyperref}

\graphicspath{{imgs/}}
\newcommand{\kk}{\mathbf{k}}
\newcommand{\ii}{\mathrm{i}}

\begin{document}

\title{Global Framework for Dynamics and Criticality of Bound States in the Continuum}

\author{Keren Wang}
\affiliation{College of Physics, Sichuan University, Chengdu 610064, China}

\author{Lujun Huang}
\email{ljhuang@phy.ecnu.edu.cn}
\affiliation{State Key Laboratory of Precision Spectroscopy, School of Physics, East China Normal University, Shanghai 200241, China}

\author{Wei Wang}
\email{w.wang@scu.edu.cn}
\affiliation{College of Physics, Sichuan University, Chengdu 610064, China}

\begin{abstract}
Bound states in the continuum (BICs) exhibit rich momentum-space dynamics, including merging, annihilation, and reconnection across symmetry directions and bands. Yet these phenomena have largely been explained case by case, without a unified framework to classify or predict them. Here we develop a global symmetry-equivariant theory for the dynamics of nondegenerate and degenerate BICs. We show that BIC dynamics can be classified into $\alpha$-, $\beta$-, and $\gamma$-processes according to root motion, which not only encompass the reported dynamics in $C_{2v}$, $C_{4v}$, and $C_{6v}$ systems but also include previously unrecognized ones. More importantly, we reveal that distinct dynamics are bridged by the criticality of BICs through a \emph{merging of merging}, in which selected direction--band branches acquire higher-order radiation zeros. Such a theory predicts which branches become critical and how to tune system parameters to realize them. Following this framework, we construct high-order BICs in full-wave calculations, realizing $Q\sim k^{-10}$ nondegenerate criticality and $Q\sim k^{-8}$ single-branch and band-paired degenerate criticalities. Our framework provides a systematic route to understanding, discovering, and selectively controlling BIC dynamics and high-$Q$ states.
\end{abstract}


\maketitle

Bound states in the continuum (BICs) are nonradiating eigenstates embedded in an open-wave continuum yet decoupled from the available radiative channels. The radiative quality factor $Q$ diverges at a BIC, while nearby quasi-BICs provide accessible ultrahigh-$Q$ resonances and enhanced light--matter interactions~\cite{Hsu2013,Hsu2016,Kang2023,Wang2024,wang_high-q_2025-1,wang_anomalous_2026}. Meanwhile, BICs appear as zeros of the outgoing far field, forming momentum-space radiation singularities whose positions, multiplicities, and connectivity organize both the high-$Q$ response and the surrounding polarization texture~\cite{Zhen2014,Yoda2020,huang_resonant_2023,yin_observation_2020,science.abj0039,wang_cavity-assisted_2026,sun_structured_2026,kang_janus_2025}. Together, these complementary properties make BICs a versatile platform for sensing, lasing, nonlinear optics, and structured-light control~\cite{Kang2023,Wang2024,sun_ultrasensitive_2026,Liu2026Upconversion,Do2026Emerging}.

The physics becomes richer when multiple BICs evolve collectively in momentum space. Under geometric tuning, their radiation zeros can move, merge, annihilate, or reconnect, reconfiguring the momentum-space high-$Q$ response~\cite{Jin2019,huang_evolution_2025,fu_high-order_2026}. Such collective transformations have been reported in a growing range of BIC systems and described by topological and perturbative theories~\cite{Jin2019,Yoda2020,Kang2021,Hu2022,Jiang2023,Gao2024,Zhang2024,ZhangLu2025,yang_analytical_2014,huang_evolution_2025,ni_three-dimensional_2024}. By reshaping the order and distribution of radiation zeros, these transformations can broaden robust ultrahigh-$Q$ responses and support nonlinear conversion, dynamic tuning, and compact lasing~\cite{Kang2022Merging,Zong2024SHG,Li2025DegenerateSHG,Ren2025Tunable,Peng2026Laser,fu_high-order_2026,hu_robust_2026,Zhou2026Tetramer,Zhang2026Monoclinic}. Yet these transformations span different symmetries, directions, and bands, and the relationships among them remain elusive. The possibility that these apparently distinct transformations are governed by a common underlying structure is particularly intriguing. This interplay is further enriched in degenerate modes, where band and momentum-space degrees of freedom become intertwined, yet their collective dynamics and transition-induced criticality remain largely unexplored~\cite{Li2025DegenerateSHG,Gao2024,Zhang2024}. It is therefore essential to develop a universal theory that describes BIC dynamics in periodic photonic systems with different types of symmetry.

Here we propose a global symmetry-equivariant framework for the two-dimensional far-field radiation map, which captures this common structure and unifies BIC dynamics and criticality across symmetry directions and bands. By organizing high-symmetry directions and, for degenerate modes, split bands into symmetry-connected branches that meet at $\Gamma$, the framework reduces the two-dimensional radiation-zero dynamics to coupled branch normal forms. Point-group equivariance then classifies the allowed root motions into $\alpha$-, $\beta$-, and $\gamma$-processes and predicts how BICs can merge, annihilate, and reconnect. The framework applies not only to nondegenerate modes but also to degenerate modes, where intertwined direction and band branches give rise to richer dynamics and criticality patterns. More importantly, transitions between distinct dynamics force selected branches to acquire higher-order radiation zeros, generating the criticality of BICs through a \emph{merging of merging}. The framework thereby predicts which dynamics are allowed, which branches become critical at their transitions, what high-$Q$ scaling laws emerge, and how system parameters can be tuned to realize them.

Guided by these predictions, we realize such critical states in full-wave calculations for both nondegenerate and degenerate modes, where selected branches acquire higher-order $Q$ scaling. For nondegenerate $C_{4v}$ modes, the predicted fold collapse selects one branch with $Q\sim k^{-10}$ while the reference branch remains at $Q\sim k^{-6}$. For degenerate modes, $C_{4v},E$ selects a single direction--band branch with $Q\sim k^{-8}$, whereas $C_{6v},E_1$ selects both directions of one band with the same scaling; the reference branches retain $Q\sim k^{-4}$. The calculations further uncover previously unrecognized dynamics and demonstrate the predicted branch selection, scaling laws, and tuning pathways, establishing symmetry as a guide for achieving directional and band-selective high-$Q$ control.

\begin{figure}[!t]
\centering
\includegraphics[width=\columnwidth]{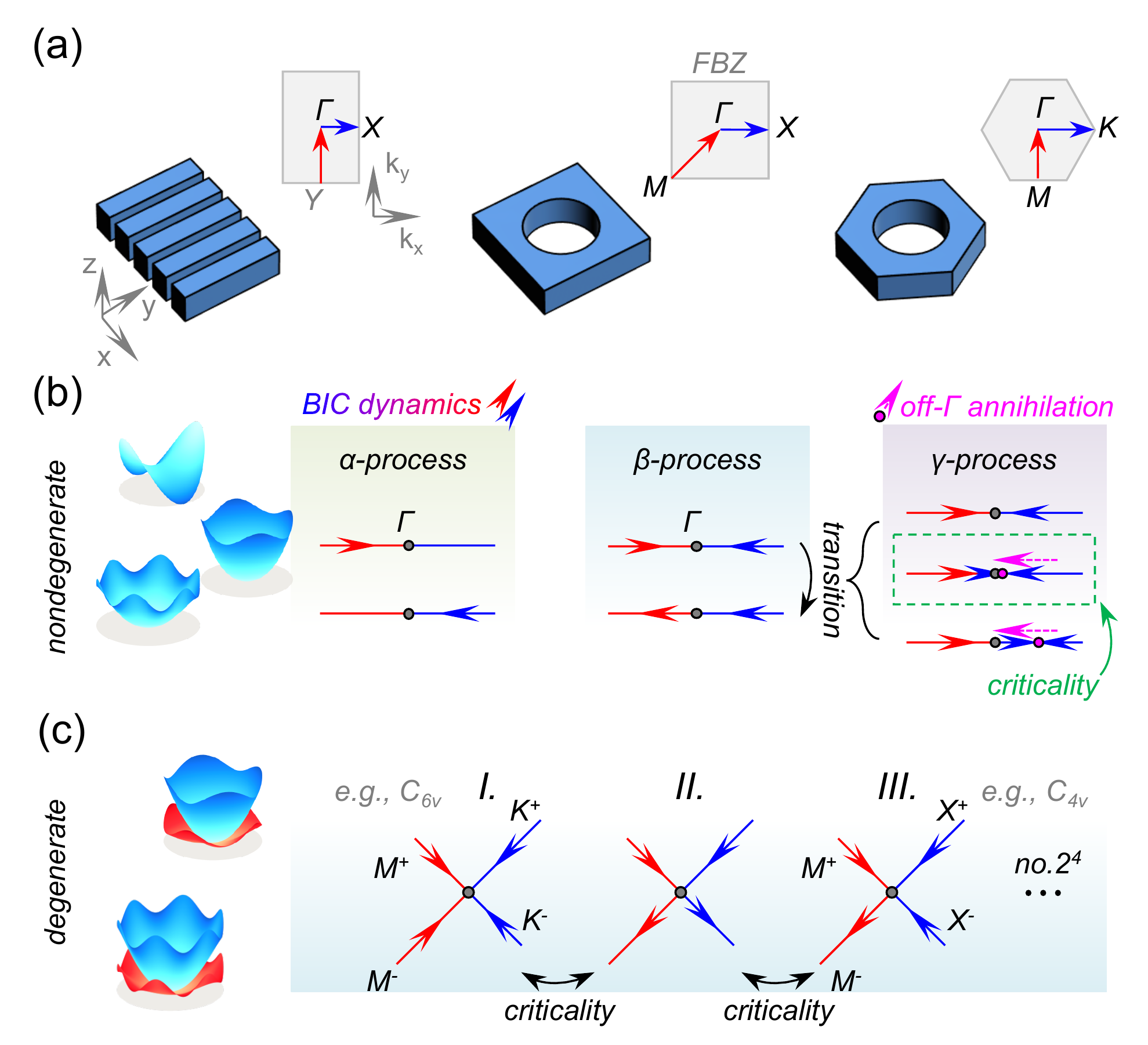}
\caption{\textbf{Symmetry-connected branches, BIC dynamics, and criticality.}
(a) Representative $C_{2v}$ grating, $C_{4v}$ square-lattice hole slab, and $C_{6v}$ hexagonal-lattice hole slab. The insets mark the two inequivalent high-symmetry branches: $Y\!\rightarrow\!\Gamma$ (red) and $\Gamma\!\rightarrow\!X$ (blue) for $C_{2v}$; $M\!\rightarrow\!\Gamma$ (red) and $\Gamma\!\rightarrow\!X$ (blue) for $C_{4v}$; and $M\!\rightarrow\!\Gamma$ (red) and $\Gamma\!\rightarrow\!K$ (blue) for $C_{6v}$.
(b) Root-flow schematics for the nondegenerate $\alpha$-, $\beta$-, and $\gamma$-processes. Magenta marks a finite-momentum pair annihilation, and the green box marks the higher-order state reached when this fold collapses into $\Gamma$.
(c) Three dynamics of four direction--band branches, ordered from left to right as four-branch annihilation (I), interband reconnection (II), and mixed three-to-one reconnection (III). The I--II transformation, realized for example in $C_{6v}$, and the II--III transformation, realized for example in $C_{4v}$, produce inequivalent criticality patterns. Representation constraints restrict the formal $2^4$ slope-sign assignments (SM).}
\label{fig:principle}
\end{figure}

To begin with, the BIC zero dynamics can be represented by a real radiation map $\bm F_\rho(\kk)=(F_x,F_y)^{\mathsf T}\in W_{\rm rad}$, obtained from the physical Jones field by the smooth port-phase alignment derived in the SM~\cite{SupplementalMaterial,Yoda2020,Hsu2017}. For a nondegenerate mode transforming according to the one-dimensional representation $\rho$, a BIC is the simultaneous zero $\bm F_\rho(\kk)=0$. We package its two real components into the complex scalar germ $\Phi_\rho(\kk)=F_x(\kk)+\ii F_y(\kk)$.
The above expression preserves both the zero condition and the radiative norm: $\Phi_\rho=0$ is equivalent to $\bm F_\rho=0$, $|\Phi_\rho|^2=F_x^2+F_y^2$, and hence $Q_{\rm rad}\propto|\Phi_\rho|^{-2}$. At the same time, this encoding converts in-plane rotations into phase multiplication, making the coefficient constraints from point-group equivariance transparent. For $g$ in the little group $G_\Gamma$,
\begin{equation}
 \bm F_\rho(g\kk)=D_{\rm rad}(g)\bm F_\rho(\kk)\rho(g)^{-1}.
 \label{eq:nondeg-equivariance}
\end{equation}
Writing $z=k_x+\ii k_y$ and $\Phi_\rho=\sum_{p,q}c_{pq}z^p\bar z^q$, a rotation $C_n$ with character $\rho(C_n)=e^{\ii\ell_\rho2\pi/n}$ implies $\Phi_\rho(C_n\kk)=e^{\ii(1-\ell_\rho)2\pi/n}\Phi_\rho(\kk)$. Comparing the transformed monomials then gives $c_{pq}\neq0$ only when $p-q\equiv1-\ell_\rho\pmod n$; mirror operations further constrain the coefficient phases and relate symmetry-equivalent directions.

Having determined the allowed two-dimensional germ, we next restrict it to an inequivalent high-symmetry direction $d$ shown in Fig.~\ref{fig:principle}(a). Along $z=ke^{\ii\theta_d}$, symmetry fixes the angular dependence, reducing the BIC condition to a radial branch amplitude. Substituting $z=ke^{\ii\theta_d}$ and $\bar z=ke^{-\ii\theta_d}$ and extracting the representation-enforced leading power gives
\begin{equation}
 \begin{aligned}
  \bm F_d(k)&=k^{m_{0,d}}g_d(s)\bm u_d,\qquad s=k^2,\\
  g_d(s)&=\alpha_d+\beta_d s+\gamma_d s^2+O(s^3),\qquad s\geq0.
 \end{aligned}
 \label{eq:normalform}
\end{equation}
Each $d$ labels an inequivalent high-symmetry direction; in the signed paths below, $k<0$ and $k>0$ place the two direction classes on opposite sides of $\Gamma$. Because every branch meets at $\Gamma$, this common junction is where roots from different directions can change their connectivity. Independent constants $\alpha_d$ allow off-$\Gamma$ roots on different branches to coalesce with the pinned $\Gamma$ zero at different control values, producing the $\alpha$-process in Fig.~\ref{fig:principle}(b). When symmetry instead locks them to a common $\alpha$, the leading roots $s_d=-\alpha/\beta_d$ are governed by the branch slopes: equal signs of $\beta_d$ give simultaneous creation or annihilation, whereas opposite signs reconnect roots through $\Gamma$, producing the two $\beta$-dynamics shown in Fig.~\ref{fig:principle}(b).

The $\alpha$- and $\beta$-processes follow from the terms through first order in $s$. Keeping the next term, $\gamma_d s^2$, permits a finite-$k$ double root at $s_{m,d}=-\beta_d/(2\gamma_d)$ and $\alpha_{m,d}=\beta_d^2/(4\gamma_d)$, describing the off-$\Gamma$ annihilation or creation of a BIC pair. Because the sign of $\beta_d$ distinguishes the neighboring $\beta$-dynamics, a continuous transformation between them forces $\beta_d$ through zero and drives this fold into $\Gamma$. As shown in Fig.~\ref{fig:principle}(b), the finite-$k$ pair event thereby coalesces with the BIC merging at $\Gamma$; we therefore call this higher-order root motion a \emph{merging of merging}. At $\alpha_d=\beta_d=0$, the selected branch satisfies $g_d\sim\gamma_d s^2$, so the fold collapse constitutes a symmetry-constrained higher-order bifurcation of the BIC root dynamics. Accordingly, we use \emph{criticality} for the resulting transition-induced higher-order state, specified by its selected branches and vanishing orders. If $g_d(s)\sim s^{j_d}$, then
\begin{equation}
 Q_d\sim k^{-2(m_{0,d}+2j_d)}.
 \label{eq:q-scaling}
\end{equation}
For a symmetry-protected nondegenerate BIC with $m_0=1$, the successive orders $j=0,1,2$ form the radiative-$Q$ scaling ladder $Q\sim k^{-2},k^{-6},k^{-10}$, respectively.

\begin{figure*}[t]
\centering
\includegraphics[width=0.98\textwidth]{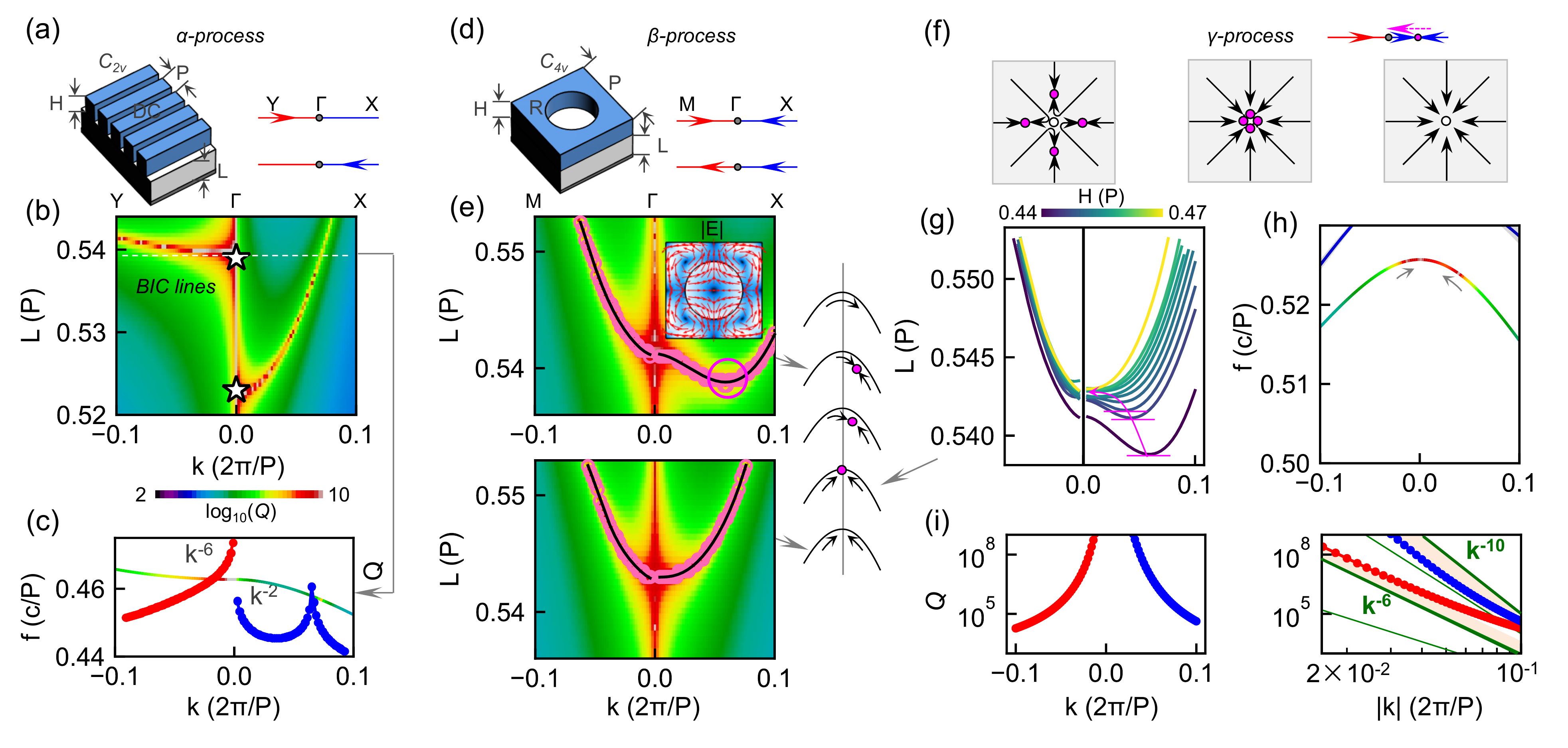}
\caption{\textbf{Nondegenerate $\alpha$-, $\beta$-, and $\gamma$-processes and the emergence of criticality.}
(a) $C_{2v},A_{1,2}$ grating-on-cavity geometry and $\alpha$-process schematic.
(b) Calculated $\log Q$ in the $(k,L)$ plane along $Y\!\rightarrow\!\Gamma\!\rightarrow\!X$. The $Y$ and $X$ roots reach $\Gamma$ separately at $L\simeq0.5400P$ and $0.5225P$ (stars).
(c) $Q$-colored dispersion and $Q(k)$ at $L=0.5395P$, showing $Q\propto k^{-6}$ on the enhanced branch and $k^{-2}$ on the other. Here $H=0.30P$ and $DC=0.5$.
(d) Square-lattice structure for nondegenerate $C_{4v}$ modes and $\beta$-process schematics.
(e) $(k,L)$ maps for cross-direction reconnection (upper) and direct annihilation (lower). Magenta markers and black fits track the BIC roots; the circled turning point is a finite-$k$ annihilation. The inset shows the near-field amplitude $|E|$.
(f) $\gamma$-process in which the finite-$k$ annihilation fold collapses into $\Gamma$.
(g) Extracted BIC curves for $0.44P\leq H\leq0.47P$, with the fold minimum reaching $k=0$.
(h) $Q$-colored band at $H=0.4580P$ and $L=0.5435P$, immediately before the BIC reaches $\Gamma$.
(i) $Q(k)$ at $H_c=0.4580P$ and $L_c=0.5430P$, shown on linear and logarithmic axes. The selected branch follows $k^{-10}$ and the reference branch $k^{-6}$, demonstrating the branch-selective criticality produced by the fold collapse. Here $R=0.30P$.}
\label{fig:nondegenerate}
\end{figure*}

\begin{figure*}[t]
\centering
\includegraphics[width=0.98\textwidth]{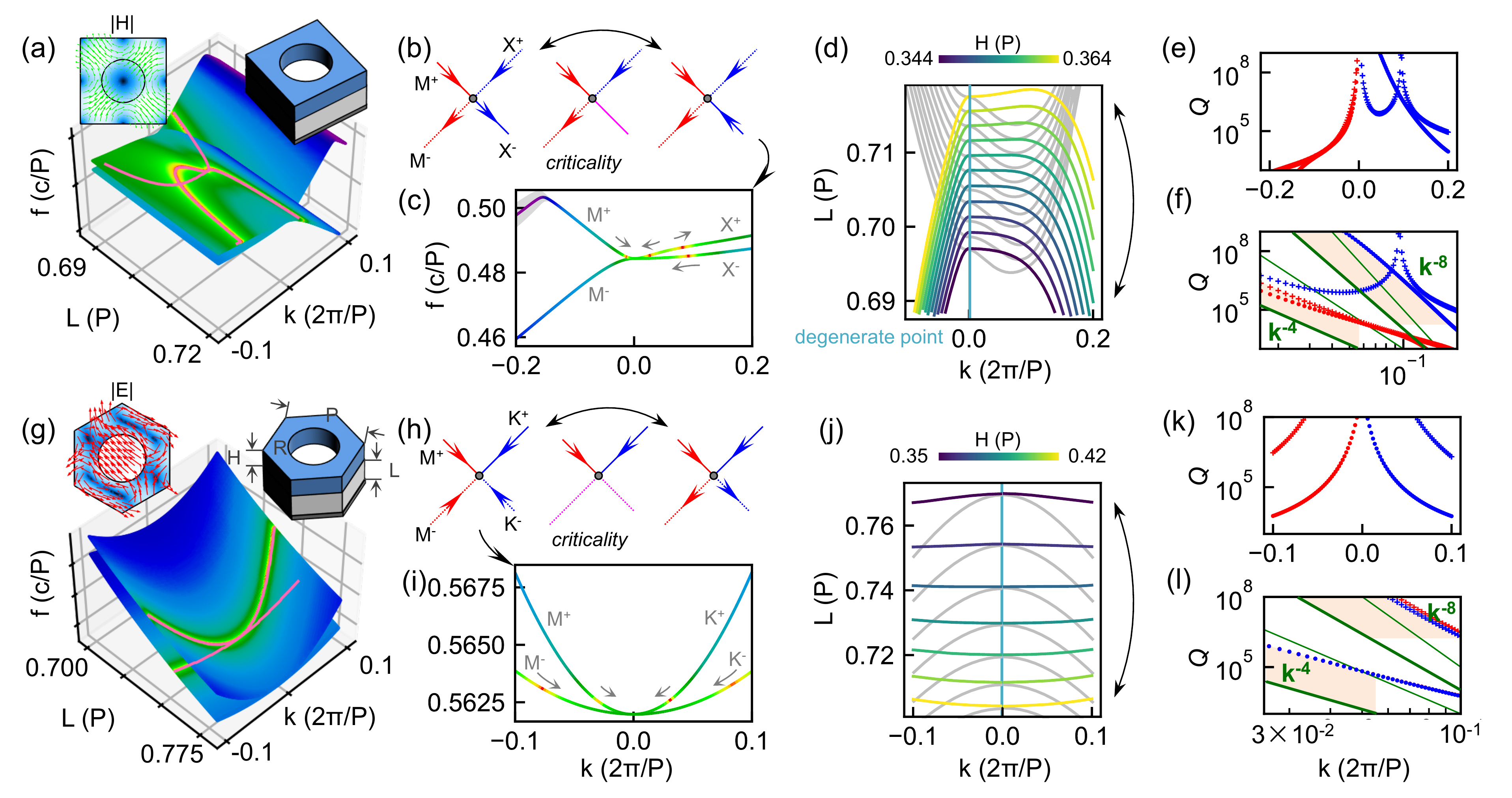}
\caption{\textbf{Representation-selected criticality in degenerate modes.}
Top row, $C_{4v},E$ single-branch criticality: (a) $Q$-colored frequency surfaces with extracted BIC trajectories; (b) transition between interband and mixed three-to-one reconnection through a single-branch higher-order BIC; (c) $Q$-colored target-band dispersion; (d) projected BIC trajectories, with the nontarget band in gray; and (e,f) linear and logarithmic $Q(k)$ curves showing the selected $k^{-8}$ branch against the $k^{-4}$ references. The critical geometry is $(H_c,L_c,R)=(0.3520,0.7055,0.25)P$.
Bottom row, $C_{6v},E_1$ band-paired criticality: (g--l) correspond respectively to (a--f), but show the transition from four-branch annihilation to interband reconnection through a band-paired higher-order BIC. Both directions of the selected band follow $k^{-8}$, while the reference branches retain $k^{-4}$. Here $(H_c,L_c,R)=(0.3750,0.73534,0.25)P$. Insets in (a,g) show the near-field amplitudes $|H|$ and $|E|$, respectively.}
\label{fig:degenerate}
\end{figure*}

For twofold-degenerate modes, the same construction requires a band projection before the branch reduction. On mirror-invariant branches, $H_{\rm eff}(\kk)\bm e_\tau=\omega_\tau\bm e_\tau$ determines the split-band eigenvectors, $\bm d_\tau=\mathcal D\bm e_\tau$ projects them onto the physical radiation channels, and port-phase alignment gives $\bm F_\tau=\mathcal R\bm e_\tau$.
As shown in Fig.~\ref{fig:principle}(c), the branch label becomes $b=(d,\tau)$, combining an inequivalent direction $d$ with a split-band index $\tau$. The signs of the four slopes $\beta_{d,\tau}$ resolve the first-order dynamics: a common sign across the branches produces four-branch creation or annihilation [Fig.~\ref{fig:principle}(c)(I)]; opposite signs between the split bands at both directions give interband reconnection through $\Gamma$ [Fig.~\ref{fig:principle}(c)(II)]; and opposite signs at only one direction give mixed three-to-one reconnection [Fig.~\ref{fig:principle}(c)(III)].

For a bright degenerate mode with $m_0=0$, tuning the allowed constant coupling $c_0$ to zero first establishes the rank-zero $\Gamma$-BIC baseline and the reference law $Q\sim k^{-4}$. Criticality then requires one additional independent condition on the branch slopes. In $C_{4v},E$, this condition can make a single slope vanish without forcing another; in $C_{6v},E_1$, symmetry ties the two directional slopes of one band, so one condition makes both vanish together. The resulting single-branch and band-paired patterns are therefore both codimension one relative to the $\Gamma$-BIC baseline, with $Q\sim k^{-8}$ on the selected branches. Together with the $c_0=0$ baseline, both realized criticalities have total codimension two from a generic bright geometry. More generally, $C_{4v},E$ allows one selected branch or an allowed pair, whereas $C_{6v},E_1$ allows only an interdirection pair, identified by full-wave continuation as one physical band; neither representation admits exactly three selected branches at this order. The complete accessibility and codimension atlas is given in the SM~\cite{SupplementalMaterial}.

Guided by the theory, we next perform full-wave eigenfrequency calculations of a Fabry--P\'erot configuration formed by a patterned dielectric layer above a spacer terminated by a perfect electric conductor. The patterned layer and spacer have refractive indices $3.48$ and $1.45$, respectively, and the period is $P=500\,\mathrm{nm}$. This configuration provides a convenient separation of structural controls: varying the spacer thickness $L$ primarily changes the constant coefficients $\alpha_b$ and therefore provides the main sweep that moves BIC roots through $\Gamma$, whereas the patterned-layer thickness $H$ and in-plane parameters such as the hole radius $R$ or duty cycle ($DC$) primarily tune the slope and higher-order coefficients $\beta_b$ and $\gamma_b$ that distinguish different dynamics. This approximate separation allows $L$ to control the root motion within a given dynamics and $H$, $R$, or $DC$ to connect distinct dynamics. Structural dimensions, root-tracking conventions, and the treatment of numerical $Q$ saturation are documented in the SM~\cite{SupplementalMaterial}.

Figure~\ref{fig:nondegenerate} organizes the nondegenerate realizations according to this $\alpha$--$\beta$--$\gamma$ hierarchy. We first consider the $C_{2v},A_{1,2}$ realization of the $\alpha$-process in Fig.~\ref{fig:nondegenerate}(a). Its orthogonal radiation channels carry independent coefficients $\alpha_Y$ and $\alpha_X$, predicting that the off-$\Gamma$ roots on the two branches coalesce with the pinned $\Gamma$ BIC at different values of $L$. The separate coalescences are visible in Fig.~\ref{fig:nondegenerate}(b), and Fig.~\ref{fig:nondegenerate}(c) identifies the resulting branch-selective anisotropic super-BIC: the enhanced branch follows $Q\sim k^{-6}$ while the other retains $Q\sim k^{-2}$~\cite{ZhangLu2025,liu_merging_2024}.

We next use the nondegenerate $C_{4v}$ structure in Fig.~\ref{fig:nondegenerate}(d) to connect the allowed $\beta$-dynamics to the nondegenerate criticality. Writing the complex momentum introduced above as $z=ke^{\ii\theta}$, equivariance restricts the representative $A_{1,2}$ germ through cubic order to
\begin{align}
 \Phi_A(\kk)
 &=c_{10}z+c_{21}z|z|^2+c_{03}\bar z^3+\cdots \notag\\
 &=z\left[c_{10}+k^2\left(c_{21}+c_{03}e^{-\ii4\theta}\right)+\cdots\right].
 \label{eq:c4a-germ}
\end{align}
The $B_{1,2}$ family has the same radial conditions after $z\leftrightarrow\bar z$ and coefficient relabeling (SM). On the two inequivalent mirror lines,
\begin{align}
 g_X(s)&=c_{10}+(c_{21}+c_{03})s+\cdots,\notag\\
 g_M(s)&=c_{10}+(c_{21}-c_{03})s+\cdots.
 \label{eq:c4a-branches}
\end{align}
Thus $\alpha_X=\alpha_M=c_{10}$, whereas $\beta_X=c_{21}+c_{03}$ and $\beta_M=c_{21}-c_{03}$ can have equal or opposite signs. Figure~\ref{fig:nondegenerate}(e) realizes both predictions at different patterned-layer thicknesses: the lower map at $H=0.4700P$ shows direct annihilation for equal signs, whereas the upper map at $H=0.4400P$ shows cross-direction reconnection for opposite signs and contains a finite-$k$ annihilation fold located by the fitted trajectory. Unlike the $C_{2v}$ anisotropic super-BIC, which is reached through a single $L$ sweep, this critical point has tuning codimension two in geometric parameter space: $c_{10}=0$ brings the BIC roots to $\Gamma$, while $c_{21}\pm c_{03}=0$ suppresses one selected branch slope. In the present realization, $L$ and $H$ implement these two independent conditions.
As $H$ is varied between these regimes, one admissible dynamics transforms into the other, so the branch slope that distinguishes them must pass through zero and drive the finite-$k$ fold into $\Gamma$. Figure~\ref{fig:nondegenerate}(f) gives the predicted fold-collapse sequence, while the extracted family in Fig.~\ref{fig:nondegenerate}(g) provides its kinematic signature: the fold minimum moves continuously to $k=0$ on approaching the critical geometry. The $Q$-colored dispersion in Fig.~\ref{fig:nondegenerate}(h) places the approaching BICs on the physical band immediately before they reach $\Gamma$. At the critical arrival of the fold at $\Gamma$, Fig.~\ref{fig:nondegenerate}(i) confirms the predicted change in radiation order, resolving $Q\sim k^{-10}$ on the selected branch against $Q\sim k^{-6}$ on the reference branch. The complete transition sequence is provided in the SM~\cite{SupplementalMaterial}.

We now extend to the four direction--band branches of a degenerate mode, whose slope relations depend on the representation. For $C_{4v},E$, symmetry gives
\begin{equation}
 \beta_{X,\pm}=c_{11}\pm c_{B1},\qquad
 \beta_{M,\pm}=c_{11}\pm c_{B2}.
 \label{eq:c4e-slopes}
\end{equation}
Because the $B_1$ and $B_2$ radiation channels are independent, one slope can vanish without imposing a partner zero on another direction or band. The theory therefore predicts codimension-one access to single-branch criticality in $C_{4v},E$; the realization shown here selects $X^-$.
Figs.~\ref{fig:degenerate}(a,c) place the extracted BIC trajectory on the $Q$-colored frequency surfaces and target-band dispersion, while Fig.~\ref{fig:degenerate}(b) sketches the transition between mixed three-to-one and interband reconnection. Figure~\ref{fig:degenerate}(d) shows the selected BIC trajectory as $H$ varies while fading the nontarget band, and Figs.~\ref{fig:degenerate}(e,f) demonstrate the $Q\sim k^{-8}$ response on $X^-$ against the $Q\sim k^{-4}$ reference branches.

By contrast, the $C_{6v},E_1$ representation with higher symmetry imposes
\begin{equation}
 \beta_{\sigma,\tau}=c_{11}+\tau c_{E2},
 \label{eq:c6e1-slopes}
\end{equation}
where $\sigma\in\{M,K\}$ and $\tau=\pm$. Because this slope depends on the band but not on the direction, the theory predicts that one scalar condition makes it vanish simultaneously on both directional branches of one band $\tau_\star$, yielding the codimension-one band-paired pattern. Figure~\ref{fig:degenerate}(g) reveals band-paired trajectories on the frequency surfaces, and Fig.~\ref{fig:degenerate}(h) shows their transition from four-branch annihilation to interband reconnection. Correspondingly, Fig.~\ref{fig:degenerate}(i) places the approaching BICs on the target band, while Fig.~\ref{fig:degenerate}(j) separates its paired motion from the nontarget band. Figs.~\ref{fig:degenerate}(k,l) then resolve $Q\sim k^{-8}$ on both selected directions against $Q\sim k^{-4}$ on the reference pair. Thus $C_{4v},E$ selects one of four branches, whereas $C_{6v},E_1$ selects two direction-related branches of one band, as predicted in the SM~\cite{SupplementalMaterial}.

In summary, our work establishes a global framework of symmetry-constrained radiation-zero dynamics. Point-group equivariance fixes the coefficient-sharing relations, branch normal forms convert them into admissible root motions, and the resulting accessibility and codimension identify which critical states can be reached with a given set of controls. The framework therefore predicts not only how BICs move and reconnect, but also which branch-selective high-$Q$ response a representation can support and how many independent tuning conditions are required to realize it.

Finally, the symmetry-connected branch atlas should be understood as an embedded skeleton of the full two-dimensional radiation map, rather than as a collection of independent one-dimensional cuts: branch normal forms along different high-symmetry directions are restrictions of a single equivariant germ and thus inherit shared, symmetry-constrained coefficients. While the skeleton theory is conveniently formulated using the phase-aligned map $\bm F(k_x,k_y)$, the full two-dimensional radiation field is the physical Jones field $\bm d(k_x,k_y)$; the alignment preserves its radiation zeros and norm~\cite{Hsu2017}. Beyond the skeleton, the same two-dimensional equivariant zero condition can also describe generic off-skeleton BICs, although their dynamics lies outside the high-symmetry branch classification developed here~\cite{Jiang2023, SupplementalMaterial}. The physical polarization phase supplies a complementary layer, for which C-point dynamics may require a separate singularity analysis~\cite{SupplementalMaterial,Hsu2017,liu_circularly_2019,sun_vertical_2026}. Extending the present admission rules to generic zeros and various polarization singularities would broaden symmetry-guided control to the full momentum-space radiation landscape~\cite{liu_complex_2026,rao_meron_2025,yin_observation_2020}.

\begin{acknowledgments}
This work was supported by the Natural Science Foundation of Sichuan Province (Grant No.~2024NSFSC0460) and the National Natural Science Foundation of China (Grant No.~12474377).
\end{acknowledgments}

\bibliography{ref}

\end{document}